# Electronic inhomogeneity and competing phases in electron-doped superconducting $Pr_{0.88}LaCe_{0.12}CuO_{4-\delta}$

Pengcheng Dai,[1,2,*] H. J. Kang,[1] H. A. Mook,[2] M. Matsuura,[2] J. W. Lynn,[3] Y. Kurita,[4] Seiki Komiya,[4] and Yoichi Ando[4]

[1]*Department of Physics and Astronomy, The University of Tennessee, Knoxville, Tennessee 37996-1200*

[2]*Condensed Matter Sciences Division, Oak Ridge National Laboratory, Oak Ridge, Tennessee 37831-6393*

[3]*NIST Center for Neutron Research, National Institute of Standards and Technology, Gaithersburg, Maryland 20899-8562*

[4]*Centeral Research Institute of Electric Power Industry, Komae, Tokyo 201-8511, Japan*

We use neutron scattering to demonstrate that electron-doped superconducting $Pr_{0.88}LaCe_{0.12}CuO_{4-\delta}$ in the underdoped regime is electronically phase separated in the ground state, showing the coexistence of a superconducting phase with a three-dimensional antiferromagnetically ordered phase and a quasi-two-dimensional spin density wave modulation. The Néel temperature of both antiferromagnetic phases decreases linearly with increasing superconducting transition temperature ($T_c$) and vanishes when optimal superconductivity is achieved. These results indicate that the electron-doped copper oxides are close to a quantum critical point, where the delicate energetic balance between different competing states leads to microscopic heterogeneity.

PACS numbers: 74.72.Jt. 75.25.+z. 75.50.Ee. 61.12.Ld

One of the most important unresolved problems in high-transition temperature (high-$T_c$) superconductors is the determination of the microscopic state when charge carriers, either holes or electrons, are introduced to the $CuO_2$ planes of their insulating long-range antiferromagnetically ordered parent compounds (*1*). One school of thought suggests that the doped charge carriers segregate into inhomogeneous patterns, such as stripes, to allow the antiferromagnetic (AF) regions to survive (*2*). In this picture, the observed quasi two-dimensional (2D) incommensurate spin density wave (SDW) in hole-doped high-$T_c$ superconductors, such as $La_{2-x}Sr_xCuO_4$ (*3*) and $La_2CuO_{4+\delta}$ (*4*), is due to remnants of the AF insulating phase that compete with superconductivity (SC) (*5,6*). Alternatively, carrier doping into the $CuO_2$ planes may evolve the system through a quantum phase transition (QPT) between competing ordered and disordered phases (*7*). While much is unknown about these competing phases and their relationship to SC, the presence of such a QPT has been reported in both hole-doped (*8-11*) and electron-doped (*12*) materials.

In this paper, we use neutron scattering to demonstrate that the AF to SC transition in electron-doped $Pr_{0.88}LaCe_{0.12}CuO_{4-\delta}$ (PLCCO) is close to a QPT (*13*) and the proximity to such a QPT leads to an inhomogeneous electronic state due to competing order parameters (*14*). Specifically, we find that when SC first appears in PLCCO with increasing doping, a quasi-2D commensurate SDW modulation is also induced, and both coexist with the three-dimensional (3D) AF state. The Néel temperature ($T_N$) of both the 2D and 3D AF phases decreases linearly with increasing $T_c$ and vanishes when optimal SC is achieved (Fig. 1f). Combined with earlier reports (*3-6,15-17*), our discovery establishes that the quasi-2D SDW modulation and its associated electronic inhomogeneity are general features of charge-carrier-doped copper oxides. Such microscopic heterogeneity naturally results from the competition of different order parameters near a QPT.

Traditionally, the transition from AF to SC in copper oxides is achieved by chemical doping, which also introduces additional complications such as structural and/or chemical disorder that may mask the nature of the transition (*13*). We choose to study this transition in electron-doped PLCCO because, as-grown, these materials are nonsuperconducting (NSC) AF insulators and SC can be achieved by only annealing the samples with minor oxygen modification that changes the charge carrier density (*18,19*). While recent neutron scattering experiments on superconducting $Nd_{1.85}Ce_{0.15}CuO_4$ (NCCO) show a drastic suppression of the static 3D AF order, how SC is established is unknown in part because the magnetic nature of the $Nd^{3+}$ in NCCO complicates the interpretation of the data (*20*). A systematic investigation on single crystals of PLCCO, where the $Pr^{3+}$ has a nonmagnetic singlet ground state (*16,17,21*), should shed new light on our understanding of the transition from the AF to SC.

We grew single crystals of PLCCO (mosaicity < 1º, cylindrical rods Ø5 mm by~20 mm) using the traveling solvent floating zone technique (*22*). As-grown, the PLCCO crystals are not superconducting and exhibit the noncollinear AF structure of undoped $Pr_2CuO_4$ (Fig. 1a) due to the weak pseudodipolar exchange interaction between Cu and Pr (*23-25*). By annealing the samples at different temperatures in pure Ar, we obtain bulk superconducting PLCCO with different $T_c$'s and therefore control the transition from the AF state to the superconducting state (*18*). Magnetic susceptibility measurements on our PLCCO crystals show the onset of bulk SC, which corresponds to zero resistivity, at $T_c$ = 24 K, 21 K, and 16 K, respectively (Fig. 1c).

We used elastic neutron scattering to probe AF order and label the momentum transfer $Q = (q_x, q_y, q_z)$ as $(H, K, L) = (q_x a/2\pi, q_y a/2\pi, q_z c/2\pi)$ in the reciprocal lattice units (rlu) appropriate for the tetragonal unit cell of PLCCO (space group *I4/mmm*, $a$ = 3.98 and $c$ = 12.27 Å). Experiments were performed in [*H, K, 0*] and [*H, H, L*] zones using triple-axis spectroscopy. In the latter geometry we also probed the 2D magnetic scattering using a two-axis energy integration mode (*26*).

The magnetic Bragg peaks can be used to determine the size and temperature dependence of the ordered Cu (and Pr, if they are significant) moments, and Fig. 2 summarizes the experiments in the [*H, K, 0*] plane for the four samples investigated. Because of the *c*-axis spin-spin coupling, the magnetic scattering of the noncollinear AF structure (Fig. 1a) is nonzero at (0.5,1.5,0), but vanishes at (0.5, 0.5, 0) (Fig. 1b) (*23-25*). The annealing process necessary for producing SC also induces a small (~0.9%) epitaxial secondary impurity phase, cubic $(Nd,Ce)_2O_3$ in the case of $Nd_{2-x}Ce_xCuO_4$ [19,27] and $(Pr,La,Ce)_2O_3$ for PLCCO [28], which has Bragg peaks at (0.5,0.5,0) and (0.5,1.5,0). For the optimally doped ($T_c$ = 24 K) sample, neither peak shows any temperature dependence, suggesting no magnetic component. Scans along (0.5,0.5,*L*) through *L* = 1, 3 where magnetic peaks are expected also found no evidence of static AF order (Figs. 3a and 3b), consistent with the phase diagram of $Pr_{1-x}LaCe_xCuO_{4-\delta}$ (*18*).

For the underdoped superconducting samples, on cooling to 5 K the intensity increases for the (0.5,1.5,0) magnetic Bragg peak (Figs. 2f and 2g) as expected. Assuming that the residual AF structure in superconducting PLCCO is identical to that of the as-grown PLCCO, we estimate that the effective static low-temperature Cu ordered moment ($M_{cu}$) reduces from 0.06 $\mu_B$ to 0.04 $\mu_B$ (with no appreciable induced Pr moment) when $T_c$ increases from 16 K to 21 K (Figs. 3d and 3f). Concomitantly, $T_N$ of the system decreases from 60 K (Fig. 3f) to 40 K (Fig. 3d). For comparison, we cut a piece off the $T_c$ = 16 K crystal and quenched SC by annealing it in air at 900 °C for 24 hours (Fig. 1c). In this NSC PLCCO sample, the $T_N$ becomes 186 K and $M_{cu}$ is 0.12 $\mu_B$ at 97 K (Fig. 3g).

In the underdoped samples, we also observed magnetic intensity at the (0.5,0.5,0) peak (Figs. 2b, 2c). The temperature dependence of the scattering at (0.5,0.5,0) and (0.5,1.5,0) for the $T_c$ = 21 K and 16 K crystals is quite similar (Figs. 3c-f), and thus suggests at first glance that they are both related to the onset of 3D AF order. However, equivalent scans in the [$H, H, L$] scattering plane indicate that the two types of peaks must have a different origin. In particular, while both (0.5,0.5,0) and (0.5,1.5,0) peaks are resolution limited in the $CuO_2$ plane (Figs. 2b, 2c, 2f, 2g), a scan along the $L$-direction of (0.5,0.5,0) in the [$H, H, L$] zone at 6 K shows only slightly enhanced "background" scattering compared to the 80 K data on the $T_c$ = 16 K crystal (Fig. 4a). Scans along the (0.5,0.5,$L$) direction find no appreciable differences between 6 K and 60 K except at (0.5,0.5,1), where the long-range 3D AF order develops (Figs. 4b and 4c). Measurements at the (0.5,0.5,1), (0.5,0.5,2), (0.5,0.5,3), and (0.5,0.5,5) positions (not shown), on the other hand, confirm quantitatively that the magnetic structure is indeed that of Fig. 1a with a $T_N \approx 60$ K and negligible induced Pr moment.

The scattering associated with the (0.5,0.5,0) peak therefore appears to be diffuse and quasi 2D along the $c$-axis direction (*29*). To test if this is indeed the case, we carried out measurements in the two-axis energy-integrated mode by aligning the outgoing wavevector $k_f$ parallel to the 2D rod direction (*26*). In this geometry, we measure the wave-vector-dependent susceptibility that includes the elastic scattering around (0.5,0.5,$L$) with 0.4 < $L$ < 0.6 (see magenta areas in Figs. 4c and 4d). The data reveal an unambiguous peak centered at (0.5,0.5, $L$) whose intensity decreases with increasing temperature (Fig. 4e). This and measurements around (0.5,0.5,0) (Fig. 2) suggest that (0.5,0.5,$L$) is weakly $L$-dependent. Therefore, underdoped PLCCO has a quasi-2D SDW modulation (Figs. 1d and 1e) that coexists with SC and 3D AF order.

The relation between $T_c$'s and $T_N$'s of our PLCCO crystals is summarized in Fig. 1f, which demonstrates that $T_N$ decreases linearly with increasing $T_c$ and vanishes when optimal superconductivity is established. While these results suggest the presence of a QPT to SC when $T_N$ approaches zero, how do we understand the coexistence of quasi-2D SDW and long-range 3D AF order and their relationship to SC?

One concern might be that the annealing process necessary for producing SC in PLCCO causes macroscopic oxygen inhomogeneities, giving rise to separate nonmagnetic SC and AF NSC phases. In this scenario one would expect to observe both a sharp $T_c$ and a robust $T_N$ weakly depending on oxygen doping as in the case of oxygen-doped $La_2CuO_4$ (*30*). However, the drastic decrease in 3D transition to long range magnetic order ($T_N$ decreasing from 186 K for NSC to 40 K for $T_c$ = 21 K sample) combined with relatively sharp magnetic transition [as in *hole*-doped $La_2CuO_4$ (*31*)] demonstrates that this is not the case. The relatively broad SC transition is then a consequence of an intrinsic electronic inhomogeneity occurring on a length scale longer than the (short) SC coherence length. While neutrons are a bulk probe and cannot determine how SC coexists with the AF phases at zero-field, magnetic field experiments show that the AF phase competes directly with SC (*32,33*). In this case, the observed commensurate quasi-2D SDW in PLCCO is analogous to the incommensurate 2D SDW in hole-doped $La_{2-x}Sr_xCuO_4$ and $La_2CuO_{4+\delta}$ (*3,4*). In the electron-doped case, the commensurate SDW may arise from the in-phase domains of "stripes" (*22*), as compared to the anti-phase domains of "stripes" in hole-doped materials (*2*). While a detailed picture of PLCCO may be obtained from future spin dynamical measurements, it appears that microscopic heterogeneity and the resulting competition of different order parameters are general features of doped copper oxides near a QPT (Fig. 1f).

This work was supported by the US NSF-DMR-0139882 and DOE under contract No. DE-AC-00OR22725 with UT/Battelle, LLC.

*E-mail address: daip@ornl.gov

Figure 1 Spin structure models, magnetic susceptibilities, and summary of results in the neutron experiments. For magnetic susceptibility, we used a 7-T SQUID magnetometer. Our neutron scattering experiments were performed on

HB-1, HB-1A, and HB-3 triple-axis spectrometers at the High-Flux Isotope Reactor (HFIR), Oak Ridge National Laboratory. Typical collimations were, proceeding from the reactor to the detector, 48'-40'-40'-120' (full-width at half-maximum), and the final neutron energy was fixed at $E_f$ = 14.78 meV with ~1 meV energy resolution. The monochromator, analyzer and filters were all pyrolytic graphite. (a) The noncollinear type-I/III spin structure of the 3D AF order in PLCCO (23,24). (b) Expected Bragg peaks from the AF structure of (a). (c) $T$-dependence of the magnetic susceptibility for the four crystals investigated. The $T_c$ = 24 K, and 21 K samples were obtained by annealing as-grown single crystals in pure Ar at 970 °C and 940 °C for 24 hours, respectively. To obtain the $T_c$ = 16 K crystal, the as-grown sample was annealed in pure Ar at 915 °C for 1 week. (d) Simplified model for checkerboard AF order in the $CuO_2$ plane, but weakly correlated along the $c$-axis. For clarity we plot the moment direction along the $c$-axis, whereas the actual spin direction is in the $a$-$b$ plane. (e) Expected scattering in reciprocal space from (d). (f) The relationship between $T_c$ and $T_N$ for samples investigated.

Figure 2 Wavevector scans in the [$H, K, 0$] zone around (0.5,0.5,0) and (0.5,1.5,0) at different temperatures for the four PLCCO samples. The in-plane and out-of-plane (vertical) resolutions of the spectrometers at (0.5,0.5,0) are ~0.025 and ~0.1-0.2 Å$^{-1}$, respectively. (a-d) Scans along the *[H, H, 0]* direction around (0.5,0.5,0) at different temperatures. (e-h) Similar scans around (0.5,1.5,0). Solid lines are Gaussian fits to the peaks on sloping backgrounds. The peaks are resolution limited and give minimum coherence length of ~200 Å.

Figure 3 (a,b) Wavevector scans through (0.5,0.5,1) and (0.5,0.5,3) for $T_c$ = 24 K PLCCO at 4.2 K. $T$-dependence of the scattering at (0.5,0.5,0) and (0.5,1.5,0) in the *[H, K, 0]* zone for (c,d) $T_c$ = 21 K, (e,f) $T_c$ = 16 K, and (g) NSC

PLCCO samples used to extract their $T_N$'s. The solid lines in (d,f) and (g) are power-law fits describing the contribution of Cu spins (*23*). The ordered Cu moments ($M_{cu}$) in (d,f) are estimated by normalizing the magnetic intensity to the weak (1,1,0) nuclear Bragg peaks without considering absorption or extinction effects. At 97 K, $M_{cu}$ = 0.12 $\mu_B$ in (g). No measurable Pr moment was induced in (d,f) while an induced Pr moment is clearly evident below 95 K in (g) [see (*24*)].

Figure 4 Measurements in the [*H,H,L*] zone on the $T_c$ = 16 K crystal. Scans along the *L*-direction around (a) (0.5,0.5,0) and (b) (0.5,0.5,1) at different temperatures. The magnetic peak in (b) is resolution limited and gives a *c*-axis ($\xi_c$) AF coherence length of 790 Å. (c) Scans along the [0.5, 0.5, *L*] direction at 6 K (closed circles) and 60 K (open circles). The peaks at *L*~0.2, 0.8 are nonmagnetic. The magenta areas in (c) and (d) mark the *L*-region probed in the two-axis integration mode. (d) The experimental geometry in the 2-axis mode in the [*H, H, L*] zone (*26*). (e) scans along the [*H,H*] direction around (0.5,0.5) with a range of *L* at different temperatures.

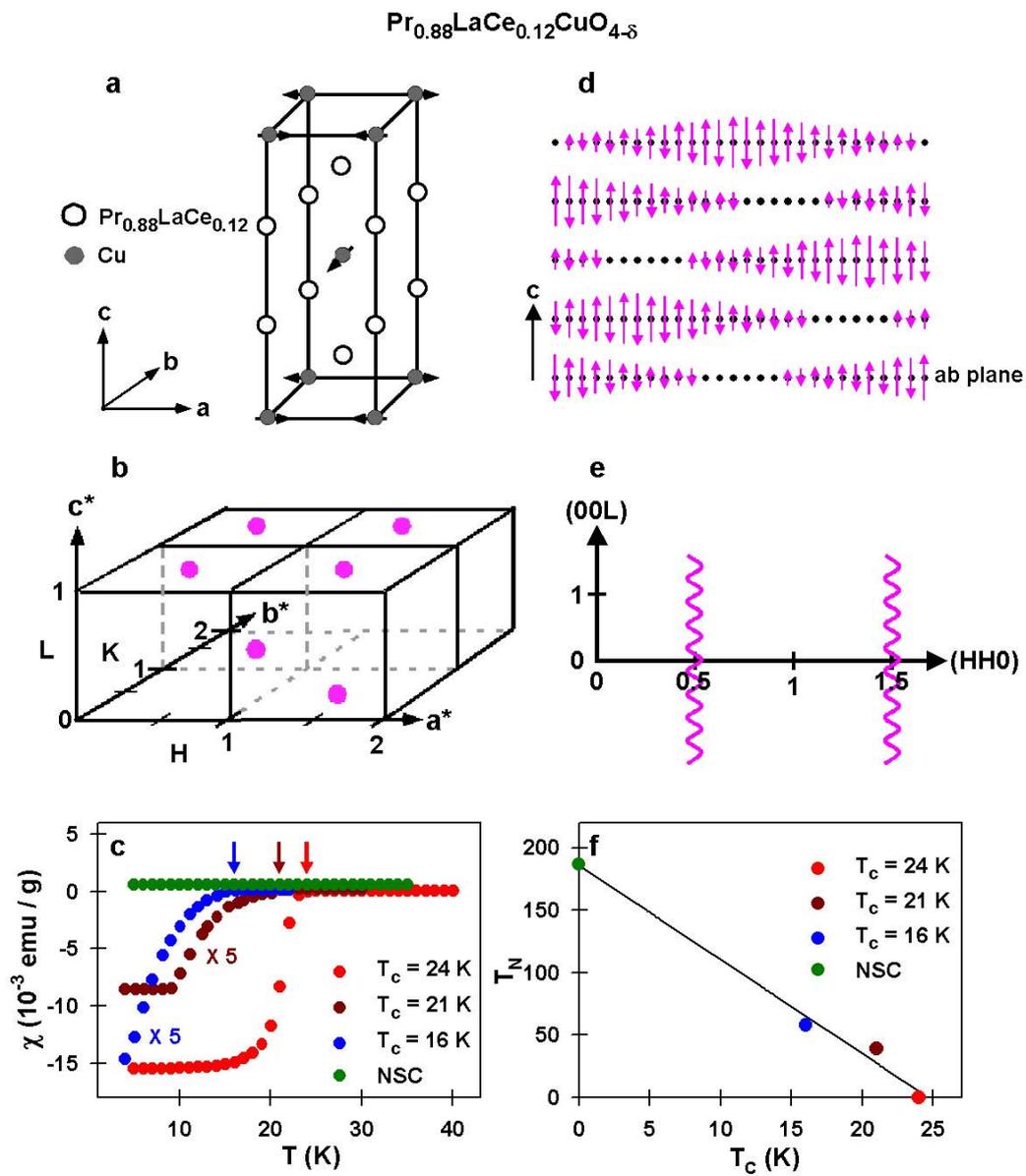

Fig. 1

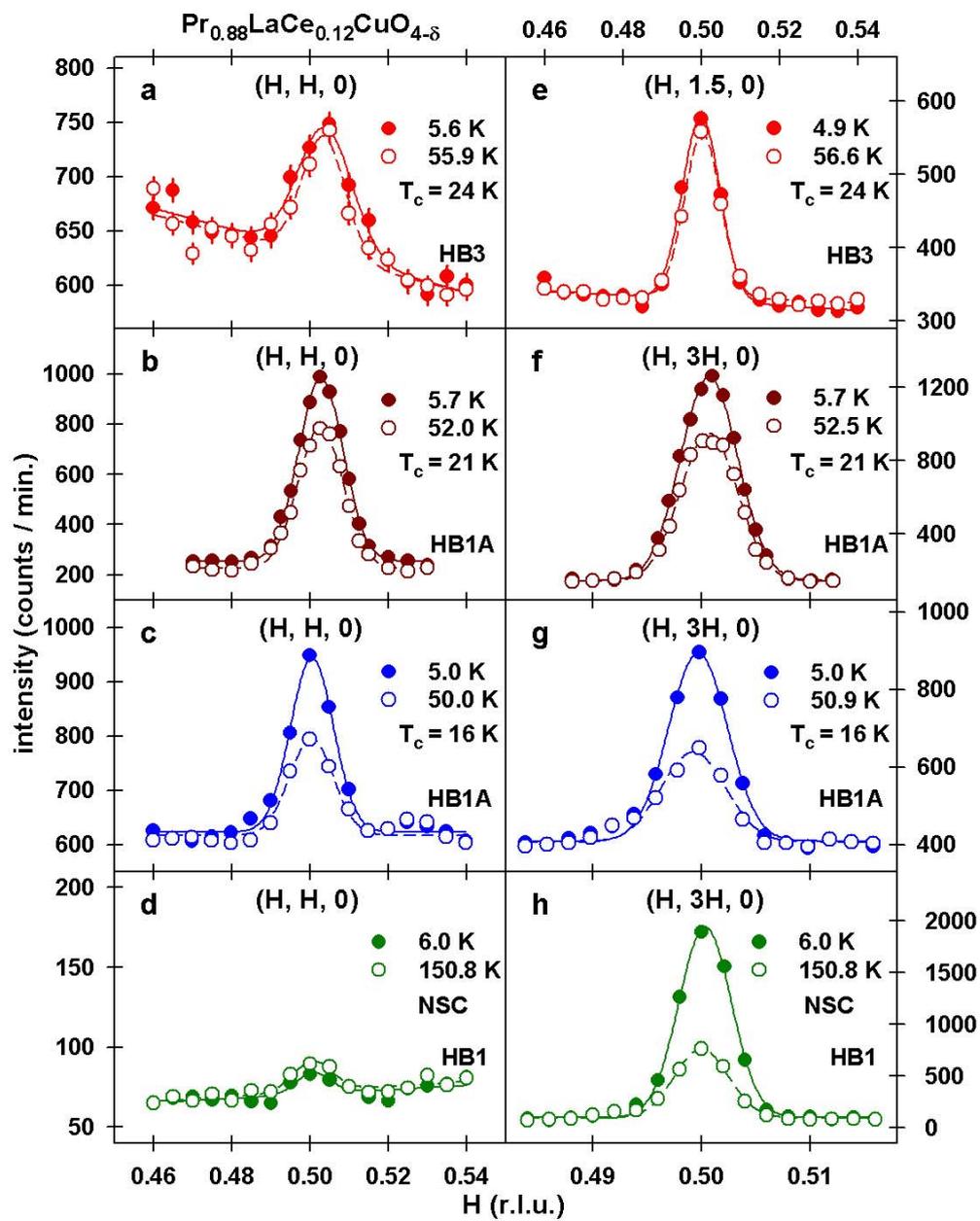

Fig. 2

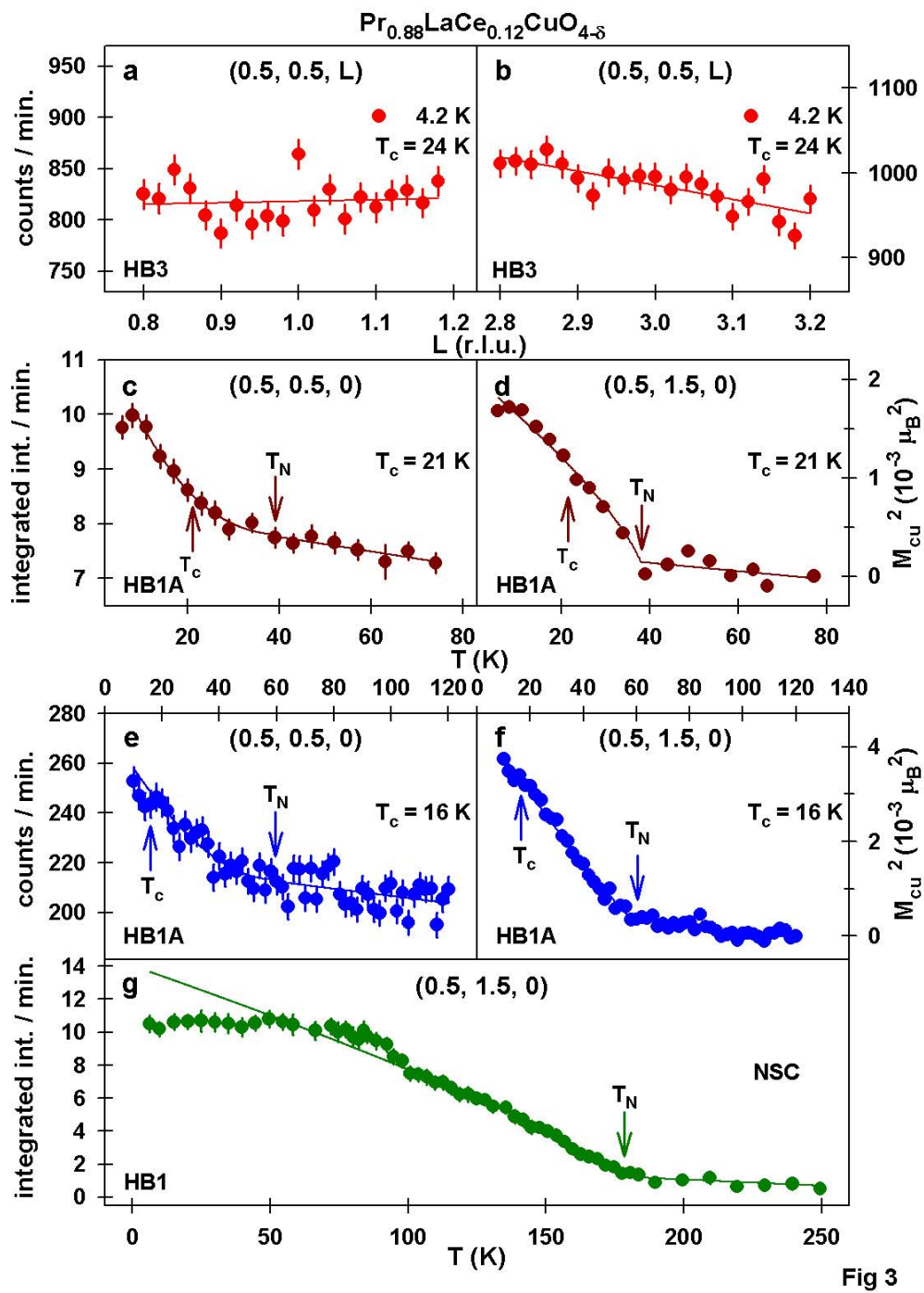

Fig 3

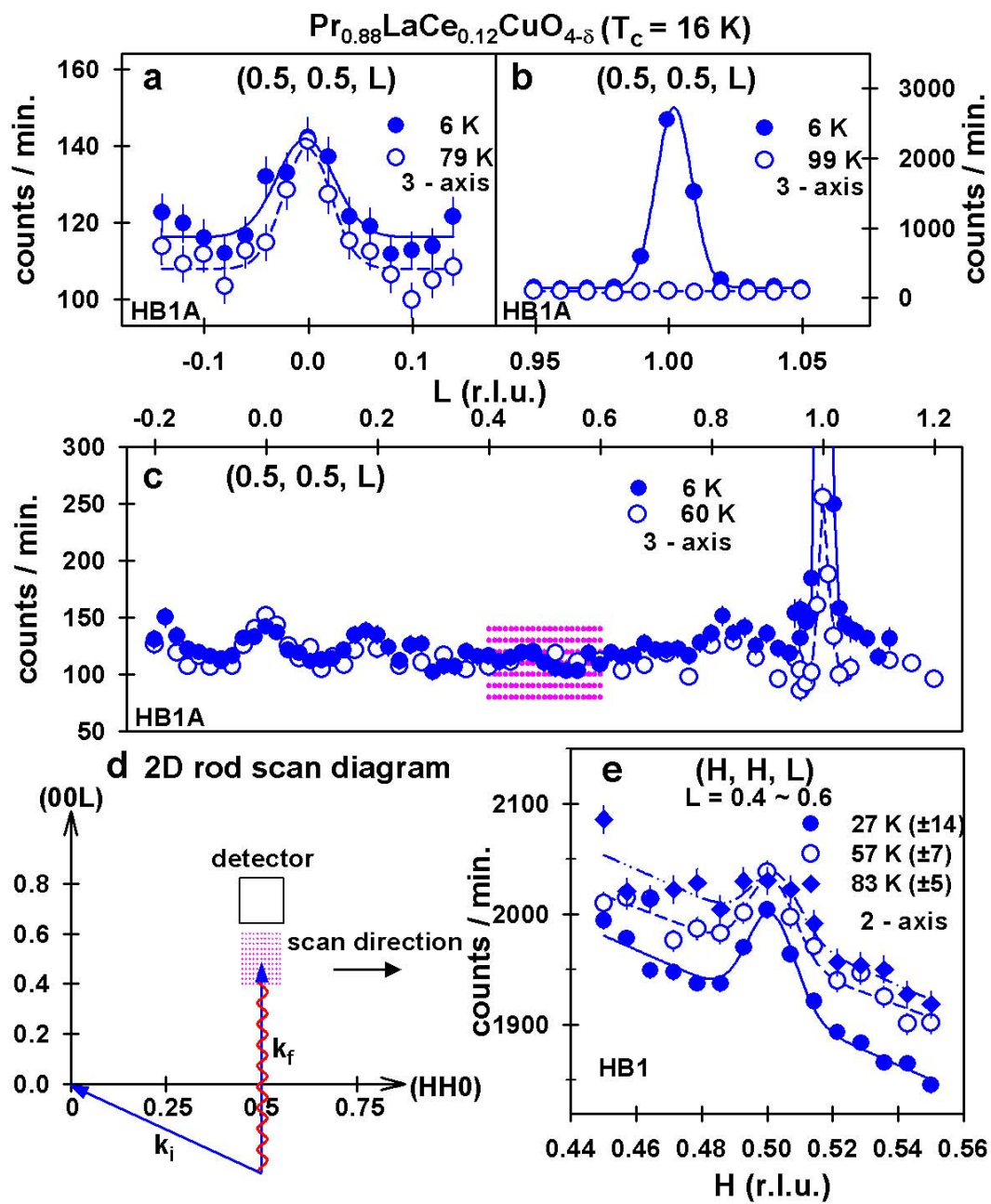

Fig. 4